# Temperature dependence of the bandgap of Eu doped {ZnCdO/ZnO}$_{30}$ multilayer structures

A. Lysak, E. Przeździecka, A. Wierzbicka, R. Jakiela, Z. Khosravizadeh, M. Szot, A. Adhikari, A. Kozanecki

Institute of Physics, Polish Academy of Sciences, al. Lotnikow 32/46, 02-668 Warsaw, Poland

*In situ* Eu-doped {ZnCdO/ZnO}$_{30}$ multilayer systems were grown on p-type Si-substrates and on quartz substrates by plasma-assisted molecular beam epitaxy. Various Eu concentrations in the samples were achieved by controlling temperature of the europium effusion cell. The properties of as-grown and annealed {ZnCdO/ZnO}$_{30}$:Eu multilayers were investigated using secondary ion mass spectrometry (SIMS) and X-ray diffraction methods. SIMS measurements showed that annealing at 700°C and 900°C practically did not change the Eu concentration and the rare earth depth profiles are uniform. It was found that the band gap depends on the concentration of Eu and it was changed by rapid thermal annealing. Varshni and Bose-Einstein equations were used to describe the temperature dependence of the band gap of {ZnCdO/ZnO}$_{30}$:Eu multilayer structures and Debye and Einstein temperatures were obtained.

## 1. Introduction

Current research in the field of II-VI oxides (such as CdO, ZnO, BeO and MgO and ternary alloys based on them) have a lot of scientific interest due to their enormous potential for a variety of optoelectronic and electronic device applications [1–3]. The relatively big energy band gap of ZnO and the band gap engineering possibilities offered by CdZnO (for narrow band gaps) and MgZnO (for wide band gaps) make the ZnO material system a promising contender in the fields of green, blue, and ultraviolet (UV) light-emitting diodes (LEDs) and lasers [4,5].

The direct band gap for ZnO is ~ 3.1 eV, while the CdO has, in addition to the direct band gap (~ 2.35 eV), two indirect band gaps ~ 1.28 and ~ 0.8 eV (at room temperature) [6,7]. It is known that the band gap of Cd$_x$Zn$_{1-x}$O ternary alloys can shift to the green region with an increase in the Cd content [8]. In this regard, the fabrication methods of Zn$_{1-x}$Cd$_x$O thin films and their properties have been intensively studied [8–10]. At the same time, the properties of heterostructures based on ZnCdO/ZnO are less studied, and the deposition of high-quality multiple quantum wells (MQW) is complicated by the difference in crystal structures of CdO and ZnO. CdO has a cubic stable phase, while ZnO crystallizes in the wurtzite phase [11]. However, the development of LEDs based on ZnCdO MQW is reported [12,13].

On the other hand, doping of oxides with rare-earth and 4d transition elements is a popular method for manipulating the optical properties of materials [14,15]. The structural and optoelectronic properties of CdO and ZnO significantly depend on the addition of lanthanides, namely samarium (Sm$^{3+}$) [16,17], cerium (Ce$^{3+}$) [18,19], yttrium (Y$^{3+}$) [20,21], lanthanum (La$^{3+}$) [22,23] neodymium (Nd$^{3+}$) [22,24] and europium (Eu$^{3+}$) [25,26]. For example, among the rare-earth ions, Eu$^{3+}$ turned out to be a promising dopant for creating light-emitting diodes (LED) in the red spectral range (620–740 nm) [27]. It is known that the ZnO doping with Eu$^{3+}$ has improved luminescent properties and retains high optical transparency [28,29]. There are several methods used to produce ZnO:Eu thin films, such as microemulsion method [27], spray pyrolysis technique [30], radio-frequency (RF) magnetron sputtering [31], pulsed laser deposition [32].

At the same time, the closeness of the ionic radius of $Eu^{3+}$ (0.95 Å) and $Cd^{2+}$ (0.97 Å) makes it easy to replace $Cd^{2+}$ ions with $Eu^{3+}$ ions in the crystal structure. As a result, the concentration of conduction electrons increases, which leads to an improvement in electrical conductivity [33]. However, the properties of the $Eu^{3+}$ doped CdO structures are still poorly examined [33–36].

Here we report the study of Eu-doped $\{ZnCdO/ZnO\}_{30}$ multilayer films grown on p-type Si and referenced transparent quartz substrates by molecular beam epitaxy (MBE). The as-deposited $\{ZnCdO/ZnO\}_{30}$ multilayer structures with Eu are annealed in oxygen atmosphere at 700°C and 900°C for 5 minutes. The as grown and the effect of post-annealing on the samples and optical properties of the multilayer systems have been studied.

## 2. Experimental details

In situ Eu-doped $\{ZnCdO/ZnO\}_{30}$ multilayer structures were grown on p-type (100) Si-substrates and referenced transparent quartz substrates by plasma-assisted (PA) MBE (Riber Compact 21). The Si p-type substrate was cut into pieces of ~ 1×1 cm. They were cleaned using buffered oxide etch for about 3 minutes, washed with distilled water, and the surfaces were purged with nitrogen gas for drying. Then p-type Si substrates degassed in a load chamber at 150°C for 1 hour. Thereafter, the substrate temperature in the growing chamber was raised to 550°C and declined to a growth temperature at 360°C (measured by thermocouple). To protect against oxidation, a thin layer of metallic Zn was deposited (by few minutes) on the Si substrate [37]. A RF cell was used to generate oxygen plasma and the RF power of the oxygen plasma was fixed at the level of 400 W and flow rate of $O_2$ gas was 3 sccm. High purity elements Zn (6 N), Cd (6 N), and Eu (4 N) were used as sources of elements from the Knudsen effusion cell. The base temperature of Cd and Zn effusion cells was fixed at 365 and 561°C, respectively. The fluxes of Cd, Zn and Eu were measured before the growth process with a beam flux monitor of the Bayard-Alpert type ionization gauge and they were ~1.33e-4 Pa and ~1.33e-5Pa for Zn and Cd, respectively. The Cd and Zn fluxes were fixed the same for all growth processes and change for Eu.

The structures of $\{ZnCdO/ZnO\}_{30}$:Eu multilayer consisted of 30 periods of ZnCdO and ZnO layers ending with a CdO cap layer. The thickness of the ZnCdO:Eu and Zn(Eu)O sublayers was controlled by the deposition time, which was 2 and 5 minutes, respectively. The CdO cap layer was deposited within 5 minutes. To achieve different concentrations of $Eu^{3+}$ in the samples "A" and "B", the europium effusion cell temperature was 100°C and 360°C, respectively. For transmittance measurements under the same conditions, the reference Eu-doped $\{ZnCdO/ZnO\}_{30}$ multilayer films were grown on quartz substrates. To study the effect of annealing on the structural and optical properties of samples, a rapid thermal processing (RTP) system (Acuthermo AW610 from Allwin21 Inc.) was applied to the film structure at 700°C and 900°C for 5 minutes in an oxygen ($O_2$) environment.

Structural studies of as-deposited and annealed Eu-doped $\{ZnCdO/ZnO\}_{30}$ multilayers were performed with a Panalytical X'Pert Pro MRD High Resolution Diffractometer. It is equipped with X-ray lamp with Cu anode and produce Cu K$\alpha_1$ X-rays ($\lambda$=1.54056 Å).

Secondary Ion Mass Spectrometry (SIMS) was employed to estimate the europium and cadmium concentration in the investigated as grown and annealed structures. The measurements were performed with a CAMECA IMS6F system. The Cs+ primary beam at the energy of 5.5 keV and the current kept at 100 nA was used. Primary beam was rastered over the area of 150 × 150 μm$^2$ and the central region of 60 μm in diameter was analyzed. Positive secondary ions $^{153}$Eu, $^{114}$Cd$^{133}$Cs and $^{16}$O$^{133}$Cs as reference were collected and counted using electron-multiplier. The Cd- or Eu-implanted ZnO layers were used as a standard. Temperature depends transmission was measured by Horiba Fluorolog-3 spectrofluorometer in wavelength range from 250 to 700 nm to estimate the band gap. Low temperature cathodoluminescence (LT-CL) investigations were performed with use of the Scanning Electron Microscope (SEM) Hitachi Su-70 equipped with Gatan MonoCL3 system and liquid helium cryostat at a temperature of ~ 5 K. The CL was excited by an incident electron beam at an accelerating voltage of 10 kV, and the beam current was 0.24 nA. The thicknesses of the as deposited samples were estimated from the cross-section SEM images and they were ~ 1.97±0.01 µm and 1.62±0.01 µm for samples "A" and "B", respectively (Fig. 1).

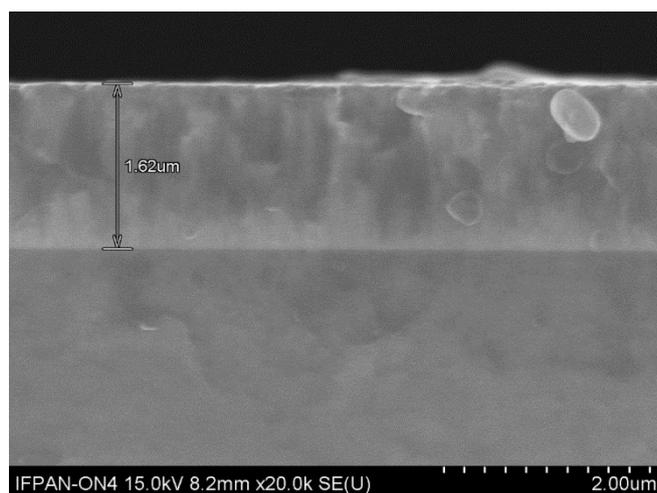

**Fig. 1.** Cross-section SEM images of the as grown sample "B."

### 3. Results and discussions

Information about the Eu presence in as grown and annealed multilayer structures was obtained using SIMS depth profiling measurements. The depth distribution of Cd, Eu, Zn and O elements is shown in Fig. 2. It can be seen, a zoom of part of the SIMS depth profile, fluctuations in Cd concentration associated with the presence of a CdO-cap layer and the individual ZnCdO and ZnO sublayers (insert in Fig. 2 d). Fluctuations in the SIMS depth profiles correspond to the high quality of as grown {ZnCdO/ZnO}$_{30}$:Eu multilayer systems and were previously observed on the SIMS profiles for the as grown {ZnO/CdO}$_m$ superlattices [38]. Relatively homogeneous Eu profile is observed in all samples. After annealing sample "B" at two temperatures (700°C and 900°C), the Cd profile becomes inhomogeneous and the cadmium concentration increases closer to the interface of the samples (Fig. 2 e, f). From the SIMS data, the concentrations of Eu atoms were recalculated using ion-implanted standards as a reference and are summarized in Table 1. The Eu concentration in the as grown {ZnCdO/ZnO}$_{30}$ multilayer structures averaged $3.6\times10^{16}$ and $9.4\times10^{18}$ atoms/cm$^3$ for samples "A" and "B", respectively. After annealing at both temperatures, the Eu concentration in the sample "A" decreases slightly. It can be seen from the Fig. 2 b,d that annealing does not induce any detectable enrichment or segregation Cd or Zn.

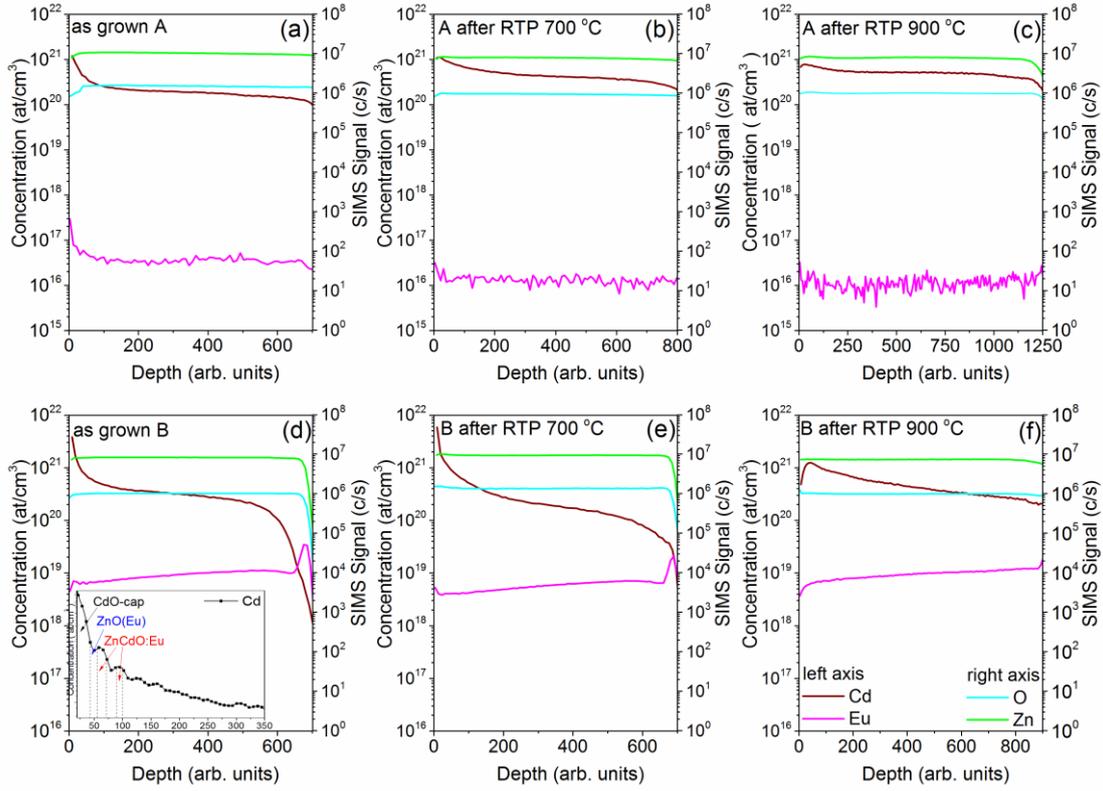

**Fig. 2.** SIMS depth profile of (a) as grown sample "A"; (b) annealed sample "A" at 700°C; (c) annealed sample "A" at 900°C; (d) as grown sample "B". The insert shows a zoom of part of the SIMS depth profile of Cd element; (e) annealed sample "B" at 700°C; (f) annealed sample "B" at 900°C.

Table 1. Structural parameters of investigated $\{ZnCdO/ZnO\}_{30}$ multilayer structure, where $d_{0002}$ is the interplanar distance, FWHM is the full width at half maximum of 0002 diffracted peak, $a$ and $c$ are the lattice constants, $\varepsilon_\parallel$ and $\varepsilon_\perp$ are the strain values as well as $n_{Eu}$ is the $Eu^{3+}$ concentration.

| Sample | | $n_{Eu}$ (at./cm$^3$) | $d_{0002}$ (Å) | FWHM (deree) | $a$ (Å) | $c$ (Å) | $\varepsilon_\parallel$ 10$^{-4}$ | $\varepsilon_\perp$ 10$^{-4}$ |
|---|---|---|---|---|---|---|---|---|
| „A" | as grown | 3.6 10$^{16}$ | 2.604 | 0.2788 | 3.247 | 5.209 | -6.2 | 7.7 |
| | RTP 700°C | 1.3 10$^{16}$ | 2.600 | 0.2357 | 3.351 | 5.200 | 4.0 | -9.1 |
| | RTP 900°C | 1.1 10$^{16}$ | 2.598 | 0.2144 | 3.258 | 5.196 | 27.7 | -17.3 |
| „B" | as grown | 9.4 10$^{18}$ | 2.603 | 0.4961 | 3.249 | 5.206 | -2.5 | 1.9 |
| | RTP 700°C | 5.8 10$^{18}$ | 2.604 | 0.2211 | 3.251 | 5.209 | 3.7 | -1.2 |
| | RTP 900°C | 8.9 10$^{18}$ | 2.602 | 0.2194 | 3.252 | 5.204 | 6.8 | -4.9 |

X-ray scan of as-grown and annealed at 700 and 900°C Eu-doped "A" and "B" multilayer systems are presented in Fig. 3. X-ray patterns showed that as grown samples "A" and "B" have characteristic peaks of orientation (10-10), (0002), (10-11), (10-12), (10-13), (0004) and (20-

22) originating from the wurtzite structure of ZnO (JCPDC 00-005-0664). The peak from the Si substrate oriented in the crystallographic direction [100] is observed in all samples (400 diffracted peak). All {ZnCdO/ZnO}$_{30}$ multilayer films exhibit strong preferential growth in the [0001] direction in wurtzite phase. The preferred growth of ZnO film with the *c*-axis perpendicular to the film plan is explained by Bouznit et. al. as well as van der Drift and associated with low surface free energy [23,39]. ZnO films doped with rare earth elements showed a tendency to growth along [0001] direction regardless of the substrate [30,40,41]. There were not found any diffraction peaks originating from Eu$_2$O$_3$ in the XRD data. This is an indication that Eu ions are probably incorporated into the ZnO host lattice.

After annealing of the samples, the position of the main 0002 diffracted peak shifts towards higher values of $2\theta$. Similar behavior was previously observed for ZnO and CdO layers doped with rare earths. It was found that upon annealing at 500-700°C in the ZnO:Eu$^{3+}$ thin films, the dominant 10-11 diffracted peak moved gradually towards higher $2\theta$ (from 31.74° to 32.04°) [31]. The shift towards higher values of the diffraction peak was observed also for ZnO:Eu films grown by the sol-gel method with Eu concentrations of 3% and 5% [42]. Edgar et al. observed a shift of the main 10-11 diffracted peak for ZnO:Sm films as a result of annealing and explained this shift by lattice relaxation [16]. Dakhel also observed a slight shift in the position of the main 111 diffracted peak of CdO, deposited on the Si substrate, towards a higher Bragg angle for CdO films doped with different Eu concentrations [33].

The *c* lattice parameter was calculated from 0002 {ZnCdO/ZnO} symmetrical reflection and *a* lattice parameter from 11-24 {ZnCdO/ZnO} asymmetrical reflection using the following formula [43]:

$$\frac{1}{d_{hkl}^2} = \frac{4}{3} \times \left(\frac{h^2 + hk + k^2}{a^2}\right) + \frac{l^2}{c^2}, \qquad (1)$$

where $d_{hkl}$ is the interplanar distance between the planes and $hkl$ is the Miller indices. The spacing of the crystal planes is determined from Bragg equation:

$$2d_{hkl} \sin \theta_{hkl} = n\lambda, \qquad (2)$$

where $\lambda$ is X-ray wavelength of the Cu K$\alpha$ line ($\lambda = 1.54056$ Å) and $\theta$ is the Bragg diffraction angle. The calculated lattice parameters $a$ and $c$ are summarized in Table 1. The $c$ values for as grown samples are larger than the lattice constants given in the standard data for ZnO ($c_0 = 5.2050$ Å) and the *a* lattice constants are smaller ($a = 3.2498$ Å) [44]. As is known, the value of the lattice constant of semiconductor materials is affected by the concentration of impurities, the presence of defects, mechanically induced strain in the lattice and different ionic radii with respect to the substituted matrix ions [22]. In our case, the increase (decrease) in these $c$ values ($a$ values) for as grown samples may be associated with the addition of atoms with a larger ionic radius such as Eu$^{3+}$ (0.95 Å) and Cd$^{2+}$ (0.97 Å), which leads to some local changes in the lattice of ZnO [41]. The presence of 2 types of not native atoms leads to the deterioration of the crystal structure of the samples. Chelouche et al. also observed an increase in the $c$ lattice parameter in ZnO films doped with Ce, and assumed that most of the Ce$^{3+}$ ions are located at substitutional sites [19]. The decrease in the $c$ values presented in Table 1 with an increase of the annealing temperature is probably due to the diffusion/migrations of Cd and Eu atoms.

Our studies allow obtaining the strain relations in our structures. The strain in direction parallel and perpendicular to the growth direction in samples were calculated (Table 1) using the expressions:

$$\varepsilon_{||} = \frac{a - a_0}{a_o}, \varepsilon_{\perp} = \frac{c - c_0}{c_o} \qquad (3)$$

where $a, c$ and $a_o, c_o$ are calculated from measurements and showed for bulk material lattice parameters of ZnO, respectively [45]. For as grown samples "A" and "B", the in-plane and out-of-plane strain have a tensile and compressive nature, respectively. After annealing at 700°C samples "A" and "B", a change in the nature of the deformation to opposite is observed.

Singh et al. observed for ZnCdO thin films, depending on the annealing temperature, that the crystallites passed from the state of tension to compression deformation in in-plane direction due to diffusion of Cd from the lattice [44].

Table 1 shows the full-width-half-maximum (FWHM) of the 0002 peaks for multilayer systems. For as grown sample "B", the FWHM is significantly higher than for sample "A". In this case, the peak broadening is related to the periodic structure of {ZnCdO/ZnO}$_{30}$:Eu multilayer structure. The decrease in FWHM with annealing temperature can be explained by the migration of Cd and Zn atoms, as a result of which the separate layers of the sample ceases to exist [38]. At an annealing temperature of 900°C, the FWHM reaches comparable values for both multilayer structures.

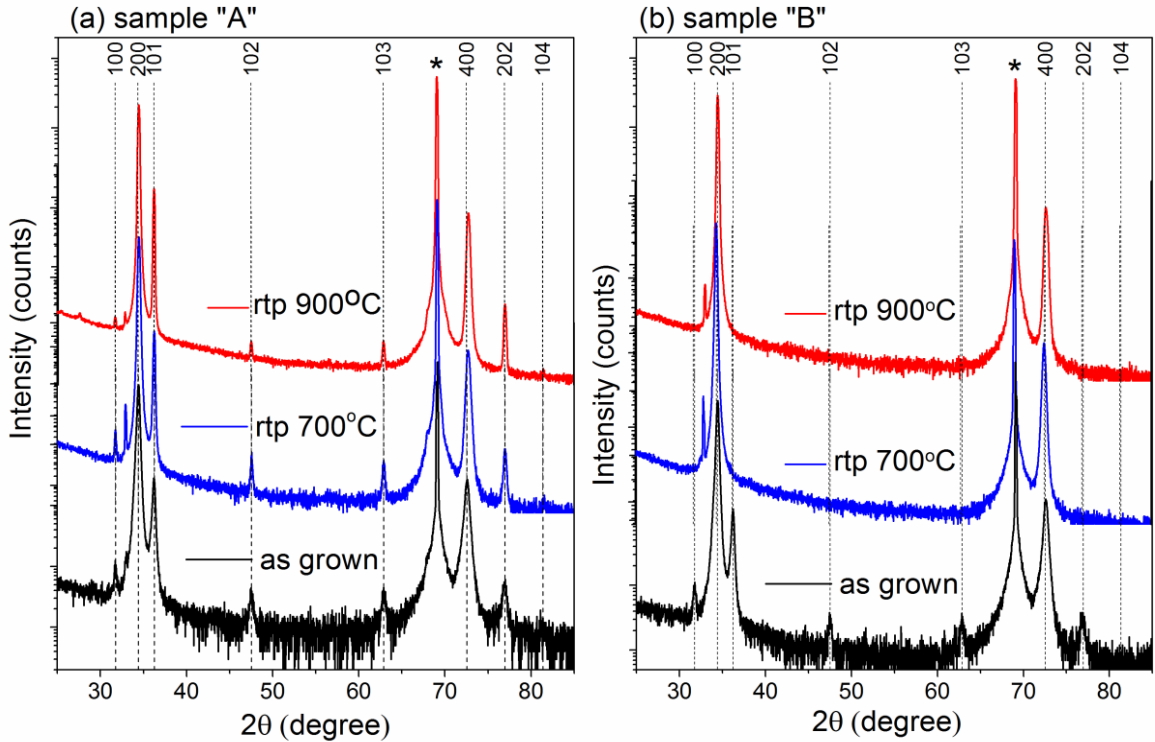

**Fig. 3.** X-ray diffraction spectra of as grown and annealed for 5 min at 700 and 900°C {ZnCdO/ZnO}$_{30}$ multilayer structure: (a) sample "A" (b) sample "B" (* indicates the peaks originated from the p-type Si (100) substrate).

To study the effect of annealing temperature on the band gap, the transmission spectra of as grown and annealed at 900°C Eu-doped {ZnCdO/ZnO}$_{30}$ structures deposited on transparent quartz substrates were measured in the temperature range of 10 - 290 K and shown in Fig. 4. As can be seen from Fig. 4, the effect of post-annealing at 900°C has a great influence on the transmission spectra of samples.

From the obtained transmission data, the absorption coefficient ($\alpha$) was calculated using the equation,

$$\alpha = -\frac{\ln T}{d}, \qquad (4)$$

where $T$ is the transmittance and $d$ is the samples thickness [46,47]. The ratio between the photon energy and the band gap ($E_g$) width is determined by the expression

$$(\alpha \cdot h\nu)^{1/n} = A(h\nu - E_g), \qquad (5)$$

where $h$ is Planck's constant, $\nu$ is the photon's frequency and $A$ is a proportionality constant. The allowed transitions dominate the basic absorption processes, giving either $n = 1/2$ or $n =$ 2 for direct or indirect transitions, respectively. The optical band gap can be obtained by extrapolation using a plot of $(\alpha \cdot h\nu)^2$ versus photon energy.

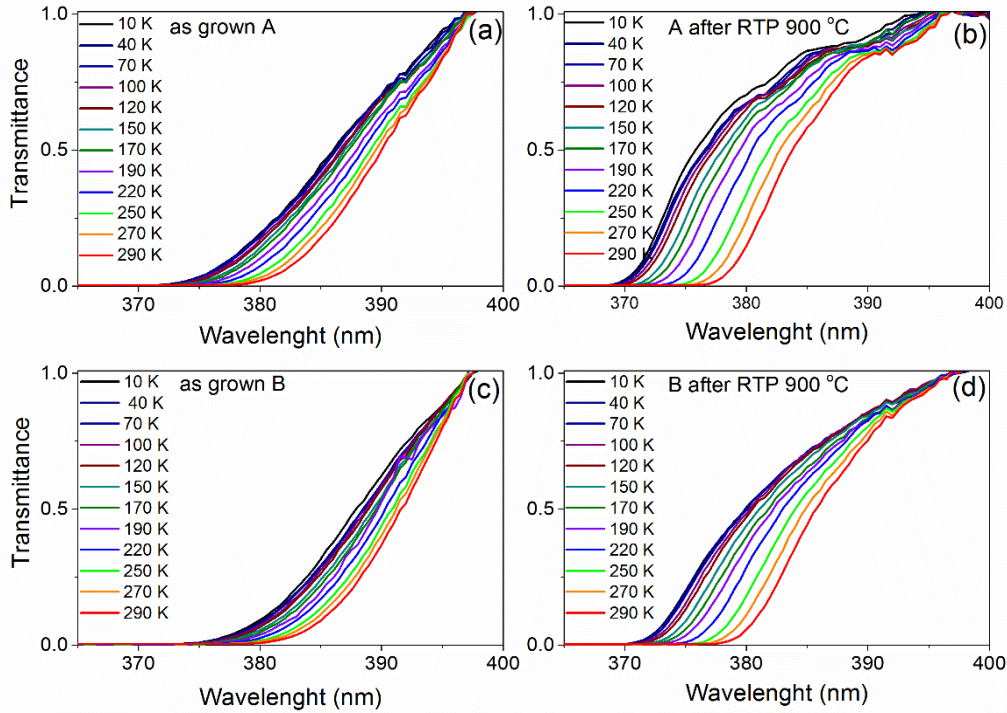

**Fig. 4.** Temperature dependent transmittance of (a) as grown sample "A" (b) annealed sample "A" at 900°C (c) as grown sample "B" (d) annealed sample "B" at 900°C.

Fig. 5 shows the temperature dependent the band gap measurements for as grown and annealed Eu-doped {ZnCdO/ZnO}$_{30}$ structures. A high Eu concentration in sample "B" leads to a shift in the band gap towards lower energies compared to sample "A". The values of optical band gaps of multilayer systems decrease with increasing temperature (Table 2). The obtained dependence of the band gap is in good agreement with the literature [31,34,48].

Table 2. Bandgap values of {ZnCdO/ZnO}$_{30}$:Eu multilayer structures; at low ($E_{10K}$) and room temperature ($E_{290K}$) and position of the CL peak measured at low temperature ($E_{CT}$).

| Sample | | $E_{10K}$ (eV) | $E_{290K}$ (eV) | $E_{CL}$ (eV) |
|---|---|---|---|---|
| „A" | as grown | 3.273 | 3.221 | 3.285 |
|  | RTP 700°C | - | - | 3.229 |
|  | RTP 900°C | 3.327 | 3.256 | 3.343 |
| „B" | as grown | 3.242 | 3.195 | 3.307 |
|  | RTP 700°C | - | - | 3.221 |
|  | RTP 900°C | 3.309 | 3.245 | 3.334 |

The band gap narrowing under high temperatures is a consequence of the intrinsic property of semiconductors at high temperatures [49]. The decrease in the band gap of Eu doped samples compared to pure ZnO is probably due to a change in the electronic structure of ZnO due to the dilution of europium in its lattice [50]. For example, the band gap decrease from 3.10 eV to 2.88 eV with an increase in the Eu$^{3+}$ concentration was observed for the CdMgZnO:xEu$^{3+}$ structure [51] and from 3.27 eV to 3.21 eV for ZnO:Eu [52]. However, the blue shift of the band gap with an increase of Eu$^{3+}$ concentration for ZnO-Eu$^{3+}$ films was observed by A. Singh et al. [41]. It is possible that a decrease in the band gap of Eu doped films is associated with an increase in oxygen vacancies [42].

F. Otieno et al. [31] found that with an increase in the annealing temperature to 900°C for the ZnO:$Eu^{3+}$, the absorption edge shifts to lower energy, which is associated with an increase in the quality of the films due to the elimination of structural defects especially oxygen vacancies [30,31]. However, after annealing of Eu in-situ doped {ZnCdO/ZnO}$_{30}$ multilayer system at 900°C, blue shift in the band gap was observed, possibly due to the decrease of the Eu and Cd concentration in the samples (Table 1). Similar behavior was observed in case of $Zn_2SiO_4$:1wt% $Eu^{3+}$ and in this case the band gap decreased linearly upon heat treatment in the range of 600 – 900°C and sharply raised at a temperature of 1000°C [53].

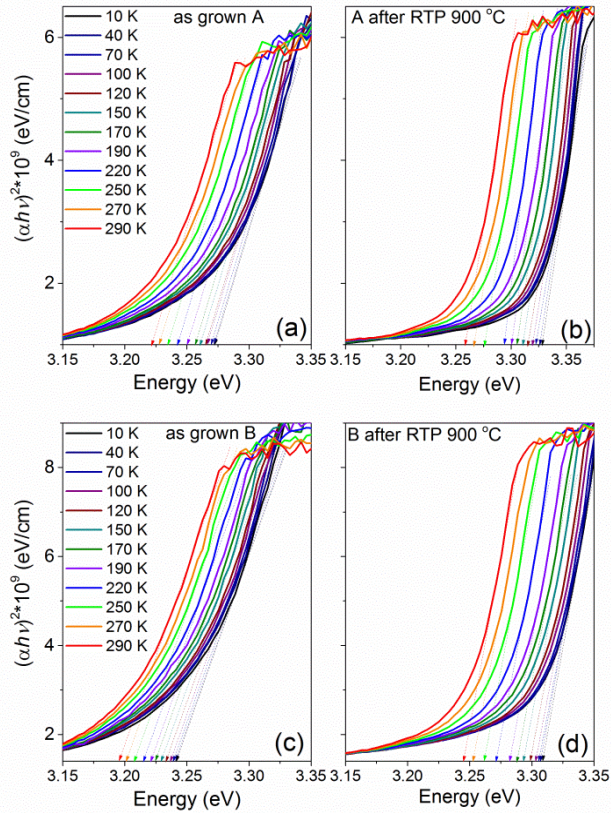

Fig. 5. Plots of $(\alpha \cdot h\nu)^2$ vs. photon energy at different temperatures of as grown and annealed Eu-doped {ZnCdO/ZnO}$_{30}$ multilayer structures on a quartz substrate.

The temperature dependence of the optical band gap of Eu-doped {ZnCdO/ZnO}$_{30}$ multilayer systems is presented in Fig. 6. Additionally, the experimental data were described by the empirical Varshni equation [49] (green lines on Fig. 6):

$$E_g(T) = E_g(0) - \frac{\alpha T^2}{T+\beta}, \quad (6)$$

where $E_g(0)$ is the optical band gap energy at 0 K, $\alpha$ and $\beta$ are fitting parameters, which are characteristic of a given semiconductor. The $\beta$ is proportional to the Debye temperature.

Another empirical relation is the Bose-Einstein (B-E) equation, which is often used to estimate the nonlinear temperature-dependent band gap shift. According to this model, the band gap energy can be determined from:

$$E_g(T) = E_g(0) - \frac{k}{exp\left(\frac{\theta}{T}\right)-1}, \quad (7)$$

where $\theta$ is Einstein characteristic temperature, which represents the average temperature of phonons interacting with the electron subsystem [54].

The obtained fitting parameters for these models are listed in Table 3. The values of the fitting parameters are similar for both "A" and "B" samples. However, it can be seen that B-E model (blue line on Fig. 6) fits better for the entire temperature range of the experimental data. Overall, both Bose-Einstein and Varshni expressions give an adequate description of the temperature dependence of the band gap. Our fitting results were presented in Table 3. As it was present both Debye's and Einstein temperatures changes due to annealing of the samples, what can be correlated with Cd and Eu diffusion in the host matrix which disturb local structure. The obtained Debye and Einstein temperatures are in agreement with data presented for MBE CdZnO layers by Wang et al. [55]. It is also comparable with the reported by He et al. [56]. Einstein temperatures reported for ZnO/ZnMgO MQWs were equal 439-519 K and 400 K for MgZnO/CdZnO MQWs [57].

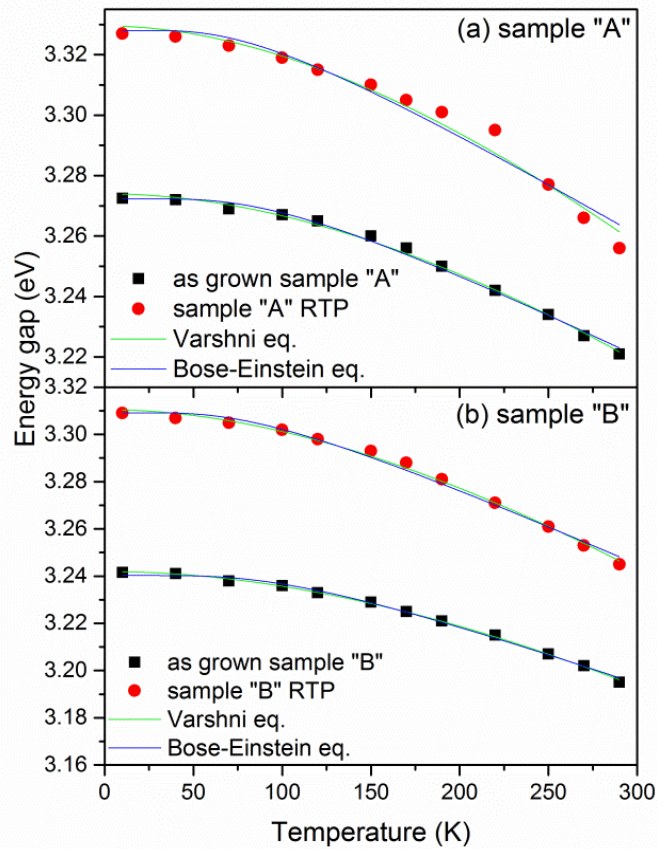

Fig. 6. The band gap energy of the Eu-doped {ZnCdO/ZnO}$_{30}$ multilayer structures as a function of temperature.

Table 3. The parameters $E_g(0)$, $\alpha$, $\beta$, $k$ and $\theta$ are obtained from the Varshni and Bose-Einstein model, respectively.

| Sample | | Varshni model | | | Bose-Einstein model | | |
|---|---|---|---|---|---|---|---|
| | | $E_g(0)$ (eV) | $\alpha\ 10^{-4}$ (eV/K) | $\beta$ (K) | $E_g(0)$ (eV) | $k$ (eV) | $\theta$ (K) |
| „A" | as grown | 3.274 | 8.2 | 1029±26 | 3.272 | 0.09 | 301 |
| | RTP 900°C | 3.329 | | 722±47 | 3.328 | | 254 |
| „B" | as grown | 3.242 | 8.2 | 1214±16 | 3.240 | 0.09 | 326 |
| | RTP 900°C | 3.301 | | 786±21 | 3.309 | | 263 |

Fig. 7 shows the normalized low-temperature cathodoluminescence (LT-CL) spectra of as grown and annealed at 700°C and 900°C Eu-doped {ZnCdO/ZnO}$_{30}$ multilayer systems deposited on p-type Si substrate. Samples exhibited ultra-violet emission at ~ 375±4 nm which correspond to near band edge radiative recombination [31,32]. The energy positions of the CL peaks are listed in the Table 2. After annealing of the multilayer structures at 700°C, a shift of the center of the UV peak is observed (from 3.285 eV to 3.229 eV and 3.307 eV to 3.221 eV, for samples "A" and "B", respectively). After annealing at 900°C, a blue shift of the CL peak for samples "A" an "B" is observed and it correlates well with the data obtained from transmittance (Table 2). L. Lee et al. observed a high-energy shift of the PL peak at room temperature after annealing (300 – 500°C) for ZnCdO samples [58].

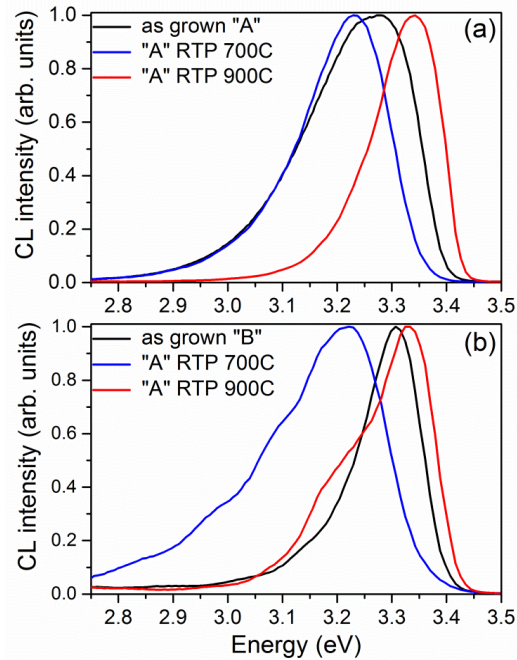

Fig. 7. Low-temperature normalized to 1 cathodoluminescence spectra of Eu-doped {ZnCdO/ZnO}$_{30}$ multilayer structures for (a) sample "A" and (b) sample "B".

Fig. 8 shows the LT-CL spectra in a wide energy range for as grown and annealed at 900°C Eu-doped {ZnCdO/ZnO}$_{30}$ multilayer structures. In the case of sample "A" the PL due to Eu and defects is not observed after annealing at 900°C (Fig. 8a). For sample "B" with a higher concentration of Eu, UV and green-blue emission are visible in the spectrum (Fig. 8b). After annealing of the sample the emission from 425 nm to 650 nm stay visible. The origin of mentioned emission in the region from 425 nm to 650 nm for annealed structures can be explained by many competing processes. One of them may be related to the deep level emission usually observed for ZnO [32,48]. The deep level emissions of ZnO are usually originated from different defect states, such as oxygen vacancies ($V_o$), zinc vacancies ($V_{Zn}$), oxygen interstitials ($O_i$) and zinc interstitials ($Zn_i$) [32,40]. On the other hand, Otieno et al. [31] detected deep-level defect green-blue emission for as grown ZnO:$Eu^{3+}$ thin films. This green-blue emission decrease and emission at ~610 nm increase due to annealing. It known, that by replacing a trivalent $Eu^{3+}$ ion with a $Zn^{2+}$ ion in the ZnO lattice, two possible types of defects can be created, such as donor centers or promoting zinc vacancies in the neighborhoods due to the charge compensation [40]. Vinoditha et al. attributed the appearance of the peak at 458 nm (blue emission) in Eu-doped ZnO nanoparticles to the recombination from the Zn interstitial defects states, which are increased due to the presence of Eu ions in ZnO crystal lattice, located below the conduction band edge [59]. Moreover, for the ZnO:Eu films, green emission at 530 nm was observed, originating from oxygen vacancies. The increase in green light emission after Eu doping was also associated with an increase in oxygen vacancies due to the replacement of Eu for Zn in the lattice [42].

Another process responsible for emission by in the visible spectrum may be the 4f-4f characteristic emission for europium. Optical transitions $^7F_0 \rightarrow {}^5D_2$, $^7F_0 \rightarrow {}^5D_1$ and $^7F_0 \rightarrow {}^5D_0$ states of $Eu^{3+}$ ions are expected at 468, 526 and 570 nm, respectively [40,52]. Due to the fact that before annealing, no emission was observed in this region **Fig. 7b** we correlated this emission with Eu doping.

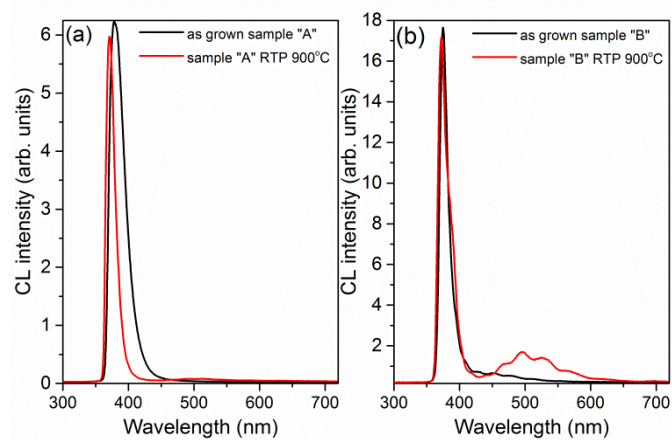

Fig. 8. Low-temperature CL spectra of as grown and annealed at 900°C (a) sample "A" and (b) sample "B".

### 4. Conclusions

{ZnCdO/ZnO}$_{30}$ multilayer structures doped with various Eu concentrations were obtained by PA-MBE method on p-type Si and quartz substrates. Analysis of SIMS data showed that annealing at 700°C and 900°C for 5 minutes in oxygen atmosphere did not lead to a significant change in the Eu concentration and its depth distribution is relatively uniform. XRD measurements showed that all multilayer systems have a predominant (0001) growth direction and the crystallographic structure of the samples did not change after RTP. Both doping of Eu and annealing affects the value of the energy gap. As a result of annealing, the energy gap of the samples increases (from 3.273 to 3.327 eV for the multilayer structure with a low Eu concentration and from 3.242 eV to 3.309 eV for the sample with high Eu concentration). The change in the band gap is attributed to migration/evaporation of Cd atoms from the samples during annealing process. The temperature dependence of the band gap for as grown and annealed {ZnCdO/ZnO}$_{30}$:Eu multilayer structures is well described by two models, namely the Varshni and Bose-Einstein. It was presented that both the Debye's and Einstein's temperatures change due to the annealing of the samples, which can be correlated with the diffusion of Cd and Eu in the host matrix, which disturbs the local structure. The annealing temperature also affected the PL signal and for high temperature of 900°C the PL peaks moved to higher energy in comparison to as-grown structures and emission in the visible range started to be observed for sample with higher Eu concentration.

**Declaration of Competing Interest**

The authors declare that they have no known competing financial interests or personal relationships that could have appeared to influence the work reported in this paper

**Data availability**

Data will be made available on request.

**Acknowledgements**

This work was supported in part by the Polish National Science Center, Grants No. 2019/35/B/ST8/01937, and 2021/41/B/ST5/00216.